\documentclass{webofc}
\usepackage[varg]{txfonts}   
\newcommand{\beq}{\begin{equation}}
\newcommand{\eeq}{\end{equation}}
\newcommand{\bea}{\begin{eqnarray}}
\newcommand{\eea}{\end{eqnarray}}
\newcommand{\dzero}{$\textrm{D}0\!\!\!\slash$}
\begin{document}
\title{Are PDFs still consistent with Tevatron data?}

\author{\firstname{Zack} \lastname{Sullivan}\inst{1}\fnsep\thanks{\email{Zack.Sullivan@IIT.edu}}}

\institute{Department of Physics, Illinois Institute of Technology, Chicago,
Illinois 60616-3793, USA}


\abstract{%
As active data taking has moved to the LHC at CERN, more and more LHC
data have been included into fits of parton distribution functions.
An anomaly has arisen where formerly excellent agreement between
theoretical predictions and experiment in single-top-quark production
at the Tevatron is no longer quite as good.  Is this indicative of a
deeper issue?
}
\maketitle
\section{Introduction}
\label{intro}

\indent\indent
In the years leading up to its discovery at the Fermilab Tevatron
\cite{Aaltonen:2009jj,Abazov:2009ii} $t$-channel single-top-quark
production played a pivotal role in the development and understanding
of improved perturbation theory and heavy quark parton distribution
functions (PDFs).  The analytic connection between this process and
deeply inelastic scattering (DIS) couples single-top to choices made
in the extraction of PDFs that imply significant constraints on higher
order corrections and scale choices.  Furthermore, large logarithms
that appear in intermediate steps of next-to-leading order (NLO) and
next-to-next-to-leading order (NNLO) calculations undergo delicate
cancellations that magnify any errors in the PDFs.  Hence, $t$-channel
single-top-quark production is one of the most constraining processes
on the consistency PDF fits at different orders, and it directly tests
the analytic framework of improved perturbation theory.

This paper introduces how modern PDF sets are failing some of the
stringent analytic tests set by $t$-channel single-top-quark
production, and speculates on some possible reasons for the failures.
In Sec.\ \ref{sec:st} I explain the constraints imposed by this
process and why it is such a stringent test.  In Sec.\
\ref{sec:failures} I demonstrate how past PDF sets were successful (or
failed in calculable ways), but current PDFs from three collaborations
CTEQ, NNPDF, and HERAPDF fail by upwards of $5\sigma$.  In Sec.\
\ref{sec:analysis} I explore possible explanations, and suggest that
we may be seeing first hints that our framework needs to be improved.
I conclude with recommendations for where to proceed to solve these
issues.

\section{$t$-channel single top, scales, and large logarithms}
\label{sec:st}

\begin{figure}[htb]
\centering
\includegraphics[width=1.5in,clip]{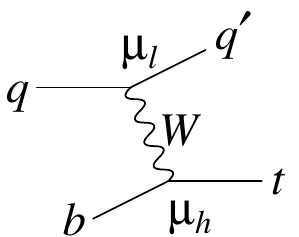}
\caption{Leading order Feynman diagram for $t$-channel
  single-top-quark production. The light-quark and heavy-quark sides
  of the diagram factorize with independent scale $\mu_l$ and $\mu_h$,
  respectively.}
\label{fig:tchan}
\end{figure}

\indent\indent
Data from the CERN Large Hadron Collider (LHC) has become precise
enough to require significantly improved theoretical calculations.
While the state-of-the-art calculations have shifted to fully
differential NNLO, few checks have been performed to confirm that the
framework and all of its pieces are both self-consistent and
consistent with all previous data.  The NNLO calculation of
$t$-channel single-top-quark production with a stable top quark
\cite{Brucherseifer:2014ama} (see Fig.\ \ref{fig:tchan} for a
leading-order figure) was recently updated to include NNLO decays of
the top quark \cite{Berger:2016oht}.  When coupled to NNLO PDFs, this
enables a comparison to experimental data at the Tevatron and LHC with
small theoretical uncertainties.

There is only one problem with the calculation in Refs.\
\cite{Brucherseifer:2014ama,Berger:2016oht}: both were performed at
fixed factorization $\mu_F$ and renormalization $\mu_R$ scales set to
the top quark mass $m_t$.  Why is this a problem?  Isn't it true that
theoretical calculations should be independent of scale choice (or at
least push the effect to higher orders)?  Not in this case.

$t$-channel single-top-quark production is a special process in that
it is analytically identical to double deeply-inelastic-scattering
(DDIS) through NLO \cite{Stelzer:1997ns,Sullivan:2004ie}.  The reason
is simply that you cannot exchange a single gluon between the
light-quark line ($q$--$q^\prime$ in Fig.\ \ref{fig:tchan}) and the
heavy-quark line ($b$--$t$ in Fig.\ \ref{fig:tchan}) and conserve
color.  In principle you could exchange two gluons at NNLO, but Refs.\
\cite{Brucherseifer:2014ama,Stelzer:1997ns,Harris:2002md} argue that
this is numerically a tiny effect, and so preserves an effective
separation between sides of the diagram.  Hence, you have DDIS with
two sets of independent scales: $\mu_l$ and $\mu_h$ for the light- and
heavy-quark lines, respectively.

This critical feature of this factorization is that when you calculate
DIS you must choose the same scale to evaluate your calculation as was
chosen to extract the PDFs from DIS data.  Otherwise, you do not get
back to the data.  This means there is a unique choice for $t$-channel
single-top: $\mu_l=Q^2$ and $\mu_h=Q^2 + m_t^2$, where $Q^2$ is the
virtuality of the $W$ boson, and the heavy-quark line sees the
heavy-quark version of DIS \cite{Stelzer:1997ns}.  A further
observation is since DIS data is used to fit the PDF, the inclusive
cross sections calculated at LO, NLO, or NNLO must all be identical
--- again you are just undoing the fit at each order.  This is a very
powerful constraint on the PDF fits.

A second important aspect of $t$-channel single-top-quark production
is related to its name.  If one tries to work in a 4-flavor scheme where
there is no $b$ quark in the initial state, a large logarithmic
divergence of the form
\beq
\alpha_s \ln \left( \frac{Q^2+m_t^2}{m_b^2} \right)
\eeq
appears at every order in the perturbative series \cite{Stelzer:1997ns}.
Numerically this term is close to 1.  Hence, you resum these large
logarithms into a $b$ PDF via the DGLAP equation \cite{Stelzer:1997ns}
\beq
b(x,\mu^2) = \frac{\alpha_s(\mu^2)}{2\pi}\ln \left(\frac{\mu^2}{m_b^2}\right) 
\int_x^1 \frac{dz}{z}
P_{bg}(z)g\left(\frac{x}{z},\mu^2\right) \,.
\eeq

This creates an improved perturbation series in which $b$ degrees of
freedom are an intrinsic part of proton structure.  In NLO and NNLO
calculations, careful subtraction of the resummed logarithms leads to
rapidly convergent series.  However, any small mistake in either the
calculation or PDF input (e.g., through faulty DGLAP evolution or poor
fits) reintroduces these large logarithms, and formerly delicate
cancellations that occur to enforce the equality of the inclusive
cross sections between orders lead to large measurable deviations.

$t$-channel single-top-quark production is one of the most sensitive
processes for testing the idea of heavy quark PDFs, the resummation
framework for improved perturbation theory, and our mapping of the
degrees of freedom of the proton onto universal parton distribution
functions.

\section{The failure of modern LO and NLO PDFs}
\label{sec:failures}

\indent\indent
Before updating the NNLO calculation to utilize the DDIS scales, I
first updated the calculation of the inclusive $t$-channel cross
section with modern PDFs using the public code ZTOP
\cite{Sullivan:2004ie}.  The ZTOP code was recently interfaced to the
new LHAPDF \cite{Buckley:2014ana} standard for accessing PDFs, and is
the standard for LO and NLO $t$-channel production.  For this study I
focus on results for the Tevatron run II at 1.96~TeV, where a combined
CDF and \dzero\ analysis exists \cite{Aaltonen:2015cra}.

In the top half of Tab.\ \ref{tab:tsig1} I show the results for
inclusive single top production at the Tevatron calculated with older
CTEQ PDFs \cite{Lai:1996mg,Lai:1999wy,Pumplin:2002vw} for which there
are no NNLO fits.  In order to be able to compare with Ref.\
\cite{Aaltonen:2015cra}, I use a top quark mass $m_t=172.5$ GeV and
the DDIS scales ($\mu_l=Q^2$, $\mu_h=Q^2+m_t^2$).  LO means a LO
matrix element, with LO PDFs, and $\alpha_s(M_Z)=0.130$.  NLO means a
NLO matrix element, with NLO PDFs, and $\alpha_s(M_Z)=0.118$.

\begin{table}
\centering
\caption{Inclusive LO and NLO $t$-channel single-top-quark cross section
($t + \bar{t}$) calculated for run II of the Tevatron ($\sqrt{S}=1.96$~TeV)
with various PDFs.}
\label{tab:tsig1}
\begin{tabular}{llll}
PDF & LO (pb) & NLO (pb) & Notes\\ \hline
CTEQ 4L/4M & 2.26   & 2.41 & 6\% deviation, known $\alpha_s$ bug  \\
CTEQ 5L/5M1 & 2.08 & 2.07 & $<0.5\%$ (bug fixed) \\
CTEQ 6L1/6M & 2.07 & 2.086 & $<0.5\%$ \\
CTEQ 6L1/6M & 1.83  & 2.086 & Scales set to $m_t$, 12\% off as expected \\ \hline
CTEQ 14 llo/nlo & 2.39 & 2.00 & 20\% LO--NLO deviation! \\
HERAPDF 1.5 lo/nlo & 1.965 & 1.798 & 9.3\% deviation! \\
HERAPDF 2.0 lo/nlo & 1.910 & 1.762 & 8.4\% deviation, NLO 12\% too small\\
NNPDF 3.0 lo/nlo & 2.33 & 2.21 & 5.4\% deviation, NLO 10\% too big\\ \hline
\end{tabular}
\end{table}

The CTEQ 4 PDF sets had a 6\% disagreement between LO and NLO, however
a bug in the running of $\alpha_s$ was discovered that accounted for
the effect.  The bug was fixed in CTEQ 5 with little change to the
data in the PDF fits, and the LO and NLO agree within the Monte Carlo
numerical integration.  Everything continued to agree with CTEQ 6, and
other distributions not shown, and the CTEQ 6 central value is
identical to the combined Tevatron fit \cite{Aaltonen:2015cra}.  One
observation from Tab.\ \ref{tab:tsig1} is that choosing the top quark
mass as a scale (as is done in Refs.\
\cite{Brucherseifer:2014ama,Berger:2016oht} and MCFM
\cite{Campbell:2004ch}) induces a large predictable 12\% shift between
orders.  While the scale dependence formally decreases at higher
orders, as it should, there is a shift with respect to the PDF
extraction by a term proportional to $\ln(Q^2+m_t^2/m_t^2)$.  The DDIS
scales are necessary here.

In the bottom half of Tab.\ \ref{tab:tsig1} I show LO and NLO results
for the inclusive cross section using modern PDFs, for which NNLO fits
exist.  Nothing agrees between LO and NLO, and the NLO calculations
between PDF sets are inconsistent.  Beginning with the latter point,
the expected 90\% confidence level NLO PDF uncertainty for this
process is $+8.8-7.3\%$ as calculated with CTEQ 14
\cite{Dulat:2015mca} PDFs --- this is consistent with error
calculations using CTEQ 6, HERAPDF, and NNPDF 3.0.  However, NNPDF
\cite{Ball:2014uwa} and HERAPDF \cite{Abramowicz:2015mha} disagree
with each other at NLO by \textit{$5\sigma$}, with CTEQ 14 splitting
the difference!  The Tevatron data \cite{Aaltonen:2015cra} is not
precise enough to rule out NNPDF and HERAPDF, but there is significant
tension with the single top data.

While the NLO discrepancies are serious, the formal question addressed
here --- the consistency between LO and NLO -- is a disaster.  The
CTEQ 14 PDFs, which are closest to the data at NLO, have a 20\%
deviation from LO when the difference should be zero.  NNPDF 3.0 is
shown as having a $5.4\%$ deviation, but this is an artifact of
choosing $m_t=172.5$ GeV.  If $m_t=175$ GeV the discrepancy also grows
to double-digits.  Even HERAPDF 1.5 and 2 are off by almost 10\%.  No
modern PDF gets back to the data.  One might be tempted to ask these
same questions at the LHC where there is more data.  Unfortunately,
all effects effects are suppressed due to accidental numerical
cancellations that occur at 13~TeV and the typical $x$ and $Q^2$
regions probed at the LHC.

\section{What is wrong with the PDFs?}
\label{sec:analysis}

\indent\indent
There are a large number of opportunities for errors to slip into any
given set of calculations: from computer code bugs to the addition of
inconsistent data.  The first place to check for problems is to check
the codes used for calculation.  ZTOP has been continuously checked
for over 13 years, and reproduces old results exactly.  One bug that
was found early in this comparison was, like CTEQ 4, LHAPDF 5
\cite{Whalley:2005nh} miscalculated $\alpha_s$ when run in multisets
mode.\footnote{Specifically, in LHAPDF 5 multiset mode, $\alpha_s$ was
  calculated for the first loaded PDF, and used for all other PDFs,
  even if the order should have been different.}  This was corrected
in LHAPDF 6.  Another observation is that there are $<0.1\%$
differences between PDFs returned from LHAPDF version of CTEQ 14, and
the CTEQ interface to CTEQ 14.  But be warned: in a NLO calculation,
which has large numerical cancellations, it can take \textit{millions}
of Monte Carlo events to converge to the same result between the
LHAPDF and CTEQ interfaces.

Another obvious possible solution to consider is that every PDF group
has been sloppy in its LO fit.  After all, the LO, NLO, and NNLO fits
are independent, and there has not been a real emphasis on LO since
2002.  This is something that needs to be examined by every group.  It
is difficult to trust the fitting procedures extended to higher orders
if we cannot reproduce known results.  However, it is logically
possible that only NNLO (and maybe NLO) PDFs are valid, and for LO we
must go back to using CTEQ 6L1.

Given that the disparities only arise in the newer PDF sets, the
obvious questions to ask are what changed between 2002 and today, and
why hasn't this been noticed before?  The answer to the second
question is that there has been a drought of LO PDFs for many years
prior to the sets listed here.  Hence, there was no opportunity to
perform these checks with the newer fitting techniques.  While
calculations and fitting methods have evolved, the biggest change has
been that the addition of jet data from the LHC, and final fits from
HERA at DESY, have begun to heavily influence the fits of proton
parton distribution functions.

Given the influx of new data sets in the fits I sought to determine
what would happen if I removed the new data.  NNPDF 3.0 includes sets
that claim to fit the PDFs without LHC data (effectively no new jet
data), or with HERA-only data.  In the first part of Tab.\
\ref{tab:altfit} I recalculate at NLO using NNPDF 3.0 sets.  The
deviations do not change when LHC data is removed --- which is
curious, since jets data \textit{should} change the NLO result; the
gluon is best measured with jets data.  A HERA-only fit by NNPDF seems
even worse, which is strange given that HERA data is DIS-like.  One
caveat is that the NNPDF fit of HERA data is not consistent with the
HERAPDF fit of HERA data shown in the second part of Tab.\
\ref{tab:altfit}, so it should probably not be used for comparison.
Unfortunately, there is no public LO version of the PDF fits with data
sets removed.  To truly make a comparison, the PDF groups need to
provide the LO complement.

\begin{table}
\centering
\caption{Inclusive LO and NLO $t$-channel single-top-quark cross section
  ($t + \bar{t}$) calculated for run II of the Tevatron ($\sqrt{S}=1.96$~TeV)
  with variations from NNPDF and HERAPDF.  Note: There are no LO variation
  sets --- making direct comparison of LO to NLO difficult.}
\label{tab:altfit}
\begin{tabular}{llll}
PDF & LO (pb) & NLO (pb) & Notes \\ \hline
NNPDF 3.0 lo/nlo & 2.33 & 2.21 & 5\% deviation\\
NNPDF 3.0 ({no LHC}) & --- & 2.22 & 5\% deviation\\
NNPDF 3.0 ({HERA}) & --- & 2.10 & 11\% deviation\\ \hline
HERAPDF 2.0 lo/nlo & 1.910 & 1.762 & 8\% deviation\\
HERAPDF 2.0 ({``JETS''}) & --- & 1.830 & $4\%$ --- $+c,\mathrm{dijets},\alpha_s$ data\\ \hline
\end{tabular}
\end{table}

\indent\indent
Turning to the HERAPDF fits, where we would have expected DIS data to
map onto DDIS data, the story is more subtle.  While the baseline
HERAPDF 2.0 fit has an 8\% deviation between orders, it does not include
all of the relevant data.  HERAPDF also performed an alternate fit
called ``JETS'' that includes the charm final state (the most closely
related to single-top), and jets data --- which technically is
differential DIS data.  Both really should be included in the fit, and
indeed the LO/NLO difference is cut in half in this case.  Still
missing is a LO version of the ``JETS'' fit.

There is a lot of room for simple mistakes to be causing the large
discrepancies between orders, with less room to explain why at most
one NLO calculation is correct.  However, we should also be asking:
is something deeper is occurring?

Hidden behind the fitting methods for proton structure is an
assumption that the degrees of freedom in the proton map to the
universal PDF shape each group uses in a clear way.  There is good
reason to believe this is not the case for the gluons that appear in
$t$-channel single-top-quark production.  The reason is that
$t$-channel single-top involves pure $V-A$ interactions at all
vertices in both production \textit{and} decay.  This is what
generates the famous angular correlations between final state
particles that was used to aid in discovery
\cite{Aaltonen:2009jj,Abazov:2009ii,Sullivan:2005ar}.

Current conservation means that only the left-handed polarization of
the gluon is probed by $t$-channel single-top-quark.\footnote{There is
  a negligible right-handed gluon contribution from the $b$ propagator
  spin-flip that goes like $(m_b/m_t)^2$.}  DIS charged-current data
that goes into PDF fits is also left-handed, though neutral-current
data has a right-handed contribution.  Jets data completely mix left-
and right-handed gluons.  Single-top sees a polarized gluon in an
unpolarized proton.  Could the numerical discrepancies we are seeing
be a consequence of ignoring spin-dependent effects in the mapping of
data onto PDF fit functions?  One numerical hint comes from another
color-singlet exchange process described at ISMD XLVI \cite{boer}:
Higgs production sees 2--5\% corrections from polarized gluons inside
unpolarized protons.  The large $\ln(m_t^2/m_b^2)$ terms which are
losing their cancellations could easily boost this effect to 10--20\%,
while in most other processes the effect would be hidden.  This is not
a proof, but it is suggestive that an effect which does have to be
there could be large enough to see.

\section{Conclusions}
\label{sec:concl}

\indent\indent
Are PDFs still consistent with Tevatron data?  Of course they are.
The $t$-channel single-top-quark data from the Tevatron is not
sufficient to place strong bounds on the $b$ quark PDF.  However, the
calculations with modern PDFs that use significant amounts of LHC data
as an input are much less in agreement than those that use older fits.
Analytic constraints between LO and NLO PDFs are no longer satisfied,
and NLO fits disagree with each other by up to $5\sigma$.  Something
is clearly wrong, the question we now need to address is why is this
failing?

There are several avenues that should be followed to resolve the
discrepancy discussed here.  LO versions of fits with alternate data
samples, such as removing LHC data or adding all HERA data, will help
point toward which data sets might be driving the broken fits.
Recalculation of the NNLO $t$-channel cross section using the DDIS
scales will allow for validation of the NNLO fits.  Will this be
enough?  It is possible the data is now so precise that we are seeing
the breakdown of the way we map gluon degrees of freedom in the proton?

How might we use $t$-channel single-top-quark production as a foil for
improving our PDFs in the future?  One idea would be to use the
analytic relationship between orders as a constraint on the fits.
More speculatively, we may wish to consider fits are specialized to
analytic constraints: e.g, fits of DIS data for DIS-like processes,
fits of jets data for jets processes, etc.  Ultimately, we should
consider whether our data is now precise enough to consider alternate
mappings of the gluon degrees of freedom in our fits that focus on
these features.  Perhaps we should be using spin-dependent PDFs even
for unpolarized cross sections.

\section*{Acknowledgments}

I would like to thank the ISMD organizers for an exceptional symposium
and environment for lively discussions.  This work was supported by the
U.S.\ Department of Energy under award No.\  DE-SC0008347.

\end{document}